\documentclass[seceq,preprint]{ptptex}

\usepackage{latexsym}

\newcommand{\nn}{\nonumber\\}

\newcommand{\ket}[1]{\left| #1 \right>}
\newcommand{\abs}[1]{\left| #1 \right|}
\newcommand{\dz}[1]{\frac{d #1}{2\pi i}}
\newcommand{\rL}{{\rm L}}

\newcommand{\Q}{Q_{\rm B}}
\newcommand{\half}{\frac{1}{2}}

\preprintnumber[3cm]{%<-- [..]: optional width of preprint # column.
hep-th/0205275\\
UT-02-29}
\title{%        %You can use \\ for explicit line-break
Open String Field Theory around Universal Solutions
}

\author{%       %Use \scshape  for the family name
Isao \textsc{Kishimoto}$^{1,}$\footnote{
E-mail: ikishimo@hep-th.phys.s.u-tokyo.ac.jp} %
and Tomohiko \textsc{Takahashi}$^{2,}$\footnote{
E-mail: tomo@asuka.phys.nara-wu.ac.jp} %
}

\inst{%         %Affiliation, neglected when [addenda] or [errata]
$^1$Department of Phyisics, University of Tokyo,
Tokyo 113-0033, Japan\\
$^2$Department of Physics, Nara Women's University,
Nara 630-8506, Japan
}

\recdate{May 27, 2002}

\abst{%         %this abstract is neglected when [addenda] or [errata]
We study the physical spectrum of cubic open string field theory around
universal solutions, which are constructed using the matter
Virasoro operators and the ghost and anti-ghost fields. We find the
cohomology of the new BRS charge around the solutions, which appear
with a ghost number that differs from that of the original theory.
Considering the gauge-unfixed string field theory, we conclude that open
string excitations perturbatively disappear after
the condensation of the string field to the solutions.
}

\begin{document}

\maketitle

%%%%%%%%%%%%%%%%%%%%%%%%%%%%%%
\section{Introduction}
%%%%%%%%%%%%%%%%%%%%%%%%%%%%%%
String field theory is a powerful tool to investigate the non-perturbative
physics of string theory, particularly 
tachyon condensation.
In cubic open string field theory (CSFT),\cite{rf:CSFT}
there is a classical solution corresponding to the tachyon
vacuum,\cite{rf:SZ-tachyon} the potential height of which is equal to the
D-brane tension and 
the physical spectrum around which contains no open string
excitations.\cite{rf:Sen,rf:Ohmori} 
At present, these studies have been carried out
by using a level truncation approximation, and the analytic form of the
tachyon vacuum solution has not yet been
obtained.\cite{rf:KS-tachyon,rf:SZ-tachyon,rf:MT,rf:ET}
It is expected that
more developments could be realized in string theory if such an
exact solution were constructed. However, there has been little progress in
the construction of the analytic
solution.\cite{rf:KP,rf:Schnabl1,rf:TT,rf:TT2,rf:KishiOhmo,rf:Kluson}

There is a universal scalar solution in CSFT that has been found
analytically and has several interesting features.\cite{rf:TT2}
It has a well-defined Fock space expression that can be written in the
universal basis, namely, the matter Virasoro operators and the ghost and
anti-ghost fields. 
Furthermore, open string excitations in the original theory are 
not physical in the theory around the solution. These properties are
necessary for the tachyon vacuum solution, since
tachyon condensation should be  a universal
phenomenon\cite{rf:SZ-tachyon,rf:SenUniv} and the D-brane should disappear
around the tachyon vacuum.\cite{rf:Sen,rf:ET,rf:HT}

In this paper, in order to explore the possibility of a universal
solution describing the tachyon vacuum, we investigate the
perturbative spectrum in the string 
field  theory around it. 
If we expand the string field around the universal solution, we obtain
the new BRS charge in the shifted theory. In order to determine the
spectrum, we solve the cohomology of the new shifted BRS
charge exactly. 

We can construct the universal solution with one parameter, for 
almost all values of which the solution is pure gauge, and the
new BRS charge can be expressed as the similarity transform of the
original BRS charge. Therefore, the cohomology of the new BRS charge has
a one-to-one correspondence with that of the original BRS charge
through the similarity transformation. However, at the boundary of the
parameter region, the similarity transformation has a sort of
singularity, and so it is impossible to reduce the cohomology of the new
BRS charge to the original one.
Consequently, we find that the non-trivial parts of the
new cohomology appear in the ghost number $0$ and $-1$ sectors only.  
Since the string field
is expanded in the ghost number one sector
in the gauge-unfixed theory, we conclude that the string field
theory around the solution contains no physical excitations
perturbatively.

This paper is organized as follows. In \S 2, we review the
universal solution in CSFT. In \S 3, we solve the cohomology of the
shifted BRS charge. Then, we show that the string theory has no physical
excitations perturbatively in \S 4.
In \S 5, we construct other possible universal solutions and find
that there is no physical spectrum around them.
We give a summary and discussion in \S 6.

%%%%%%%%%%%%%%%%%%%%%%%%%%%%%%
\section{Universal solution}
%%%%%%%%%%%%%%%%%%%%%%%%%%%%%%

The action in cubic open string field theory is given by
\begin{eqnarray}
 S[\Psi] = \frac{1}{g}\,\int \left( \Psi*\Q \Psi 
+ \frac{2}{3}\,\Psi*\Psi*\Psi\right),
\end{eqnarray}
where $\Psi$ is the string field and $\Q$ is the Kato-Ogawa BRS charge.
The equation of motion is
\begin{eqnarray}
 \Q\Psi+\Psi*\Psi=0.
\end{eqnarray}
One of the solutions of this equation has been constructed as\cite{rf:TT2}
\begin{eqnarray}
\label{Eq:solution}
 \Psi_0(h) = Q_\rL\left(e^h-1\right) I 
             -C_\rL\left((\partial h)^2 e^h \right) I,
\end{eqnarray}
where $I$ denotes the identity string field, and $Q_\rL$ and $C_\rL$ are
defined by
\begin{eqnarray}
 Q_\rL(f) = \int_{C_\rL} \dz{w} f(w) J_{\rm B}(w),\ \ \ 
 C_\rL(f) = \int_{C_\rL} \dz{w} f(w) c(w),
\end{eqnarray}
where $J_{\rm B}(w)$ and $c(w)$ are the BRS current and the ghost field,
respectively. 
The contour $C_\rL$ denotes the path along the left-half of 
strings. 
The function $h(w)$ satisfies $h(-1/w)=h(w)$ and $h(\pm i)=0$, and
as $w$ approaches $\pm i$, $(\partial h)^2 e^h$ must approach zero
sufficiently fast that $(\partial h)^2 e^h$ cancels 
the midpoint singularity of the ghost field on the identity and hence
that the solution has
a well-defined Fock space expression.

The BRS current is written in terms of  the matter Virasoro operators and the
ghost and anti-ghost fields, and the identity string field is
represented by the total Virasoro operators.\cite{rf:RZ}
Therefore, this  solution has a universal expression for arbitrary
background.

If we expand the string field around the classical solution as
\begin{eqnarray}
 \Psi=\Psi_0 + \Psi',
\end{eqnarray}
the action becomes
\begin{eqnarray}
\label{Eq:shiftaction}
 S[\Psi_0+\Psi'] = S[\Psi_0] +
 \frac{1}{g}\,\int \left( \Psi'*\Q' \Psi' 
+ \frac{2}{3}\Psi'*\Psi'*\Psi'\right).
\end{eqnarray}
The first term on the right-hand side corresponds to the potential
height of the solution, but we do not discuss it in this paper.
The shifted BRS charge is given by
\begin{eqnarray}
\label{Eq:Qh}
 \Q' = Q(e^h) -C\left((\partial h)^2 e^h\right),
\end{eqnarray}
where we have defined
\begin{eqnarray}
 Q(f) = \oint \dz{w}\,f(w) J_{\rm B}(w),\ \ \ 
 C(f) = \oint \dz{w}\,f(w) c(w).
\end{eqnarray}
In general, it turns out that if the BRS charge is written in the form
 $Q(f)+C(g)$, it must be expressed as $\pm \Q'$
or $C(g)$ because of its nilpotency.\footnote{
This follows from the relations $\{Q(f),Q(g)\}=2\{Q_B,C(\partial f\partial g)\},\{Q(f),C(g)\}=\{Q_B,C(fg)\}.$
}

In the shifted theory around the solution, let us consider the
redefinition of the string field as 
\begin{eqnarray}
\label{Eq:redef}
 \Psi'' = e^{-q(h)} \Psi',
\end{eqnarray}
where, using the ghost number current $J_{\rm gh}$,
$q(f)$ is defined by
\begin{eqnarray}
 q(f) = \oint \dz{w} f(w) J_{\rm gh}(w).
\end{eqnarray}
Through the redefinition given in Eq.~(\ref{Eq:redef}), the shifted BRS charge is transformed into the
original one,
\begin{eqnarray}
 e^{-q(h)}\Q' e^{q(h)} = \Q.
\end{eqnarray}
Then, the action in Eq.~(\ref{Eq:shiftaction}) is transformed into the
action with the original BRS charge.
This fact suggests that the solution might be pure gauge.

However, we can obtain non-trivial solutions if the string field
redefinition itself is ill-defined. The operator $q(h)$ is separated as
\begin{eqnarray}
 q(h) = q_0(h)+q^{(+)}(h)+q^{(-)}(h),
\end{eqnarray}
where $q_0(h)$, $q^{(+)}(h)$ and $q^{(-)}(h)$ correspond to the zero
mode part and the positive and negative frequency parts of $q(h)$, respectively.
The operator $e^{\pm q(h)}$ in the redefinition can be rewritten in `normal
ordered form' as
\begin{eqnarray}
 e^{q(h)} = \exp\left(\half [
 q^{(+)}(h),\,q^{(-)}(h)]\right) 
  e^{q_0(h)}e^{q^{(-)}(h)}e^{q^{(+)}(h)}.
\end{eqnarray}
If the commutator of $q^{(+)}(h)$ and $q^{(-)}(h)$ becomes singular, the
operator $e^{q(h)}$ becomes ill-defined, and so we cannot redefine the string
field as in Eq.~(\ref{Eq:redef}). In this case, a non-trivial solution
can be found for such a function $h(w)$.

In the following, we consider the classical solution for the specific
function
\begin{eqnarray}
\label{Eq:ha}
 h_a(w) = \log\left(1+\frac{a}{2}\left(w+\frac{1}{w}\right)^2\right).
\end{eqnarray}
This function $h_a(w)$ can be written in a Laurent expansion as
\begin{eqnarray}
\label{Eq:hexp}
 h_a(w) &=& -\log(1-Z(a))^2 -\sum_{n=1}^\infty \frac{(-1)^n}{n}
Z(a)^n
\left(w^{2n}+w^{-2n}\right),\nn
 Z(a) &=& \frac{1+a-\sqrt{1+2a}}{a},
\end{eqnarray}
where $-1\leq Z(a)<1$ ($-1/2\leq a<\infty$). 
This series is convergent within the annulus $\abs{Z(a)}^{1/2}<
\abs{w}<1/\abs{Z(a)}^{1/2}$. 
For this $h_a(w)$, the commutator of $q^{(\pm)}$ can be
evaluated as
\begin{eqnarray}
 [q^{(+)}(h_a),\,q^{(-)}(h_a)] = 2\sum_{n=1}^\infty \frac{1}{n}Z(a)^{2n}
 =-2\log(1-Z(a)^2).
\end{eqnarray}
In the case $Z(a=-1/2)=-1$, the commutator becomes divergent, and the
string field redefinition is ill-defined. Then, we can obtain a
non-trivial universal solution. It should be noted that for $a=-1/2$,
the Laurent expansion of Eq.~(\ref{Eq:hexp}) is ill-defined, because the
annulus $\abs{Z(a)}^{1/2}< \abs{w}<1/\abs{Z(a)}^{1/2}$ shrinks to a
unit circle. However, the expansion becomes well-defined on the unit
circle as a Fourier expansion if we redefine $w$ as $w=\exp(i\sigma)$. 
It is sufficient for our argument to expand $h(w)$ in
a Fourier series.

For $a=-1/2$ case, the non-trivial universal solution is given
by\cite{rf:TT2} 
\begin{eqnarray}
\label{Eq:solution2}
 \Psi_0 = Q_\rL\left(-\frac{1}{4}\left(w+\frac{1}{w}\right)^2\right) I
 +C_\rL\left(w^{-2}\left(w+\frac{1}{w}\right)^2\right) I,
\end{eqnarray}
and the shifted BRS charge becomes
\begin{eqnarray}
\label{Eq:shiftedBRS}
 \Q'&=& Q\left(-\frac{1}{4}\left(w-\frac{1}{w}\right)^2\right)
   +C\left(w^{-2}\left(w+\frac{1}{w}\right)^2\right).
\end{eqnarray}

%%%%%%%%%%%%%%%%%%%%%%%%%%%%%%
\section{Cohomology of the new BRS charge}
%%%%%%%%%%%%%%%%%%%%%%%%%%%%%%

The shifted BRS charge contains several level operators, and therefore it
is difficult to solve its cohomology in the usual manner. We
consider rewriting the shifted BRS charge as a fixed level expression
through some transformation.
From the operator product expansions of the ghost number current
with the BRS current and the ghost field, we can derive the following
commutation
relations\cite{rf:TT2}:
\begin{eqnarray}
\label{Eq:qQ}
 \left[q(f),\,Q(g)\right] &=& Q(fg) -2 C(\partial f \partial g), \\
\label{Eq:qC}
 \left[q(f),\,C(g)\right] &=& C(fg).
\end{eqnarray}
Using the commutation relations in Eqs.~(\ref{Eq:qQ}) and (\ref{Eq:qC}),
we find that
\begin{eqnarray}
\label{Eq:EqQ}
 e^{q(f)} Q(g) e^{-q(f)} &=&
 Q(g\,e^f) - C\left(\{(\partial f)^2 g +2 \partial f
  \partial g\}e^f\right), \\
\label{Eq:EqC}
 e^{q(f)} C(g) e^{-q(f)} &=& C(g\,e^f).
\end{eqnarray}
Then, the shifted BRS charge of Eq.~(\ref{Eq:Qh}) is transformed as
\begin{eqnarray}
\label{Eq:EqQd}
 e^{q(f)} \Q' e^{-q(f)} &=& Q(e^{h+f})-C(\{\partial(h+f)\}^2 e^{h+f}).
\end{eqnarray}
Applying Eq.~(\ref{Eq:EqQd}) to
Eq.~(\ref{Eq:shiftedBRS}), we find that the  
shifted BRS charge for $a=-1/2$
can be transformed into a fixed level operator as 
\begin{eqnarray}
\label{Eq:Qtwo}
 e^{q(\lambda)}\, \Q' e^{-q(\lambda)}
 = -\frac{1}{4}Q(w^2)+C(1),
\end{eqnarray}
where $\lambda(w)$ is given by
\begin{eqnarray}
 \lambda(w) = 2 \sum_{n=1}^\infty \frac{1}{n}w^{-2n} 
            = -2\log(1-w^{-2}).
\end{eqnarray}
Using the oscillator expressions, these equations can be written as
\begin{eqnarray}
\label{Eq:qQq}
&& e^{q(\lambda)}\Q' e^{-q(\lambda)} = -\frac{1}{4}Q_2 +c_2, \\
&& \Q' = \frac{1}{2}\Q-\frac{1}{4}(Q_2+Q_{-2})
 +2 c_0 +c_2+c_{-2} ,\\
&& q(\lambda) = 2\sum_{n=1}^\infty \frac{1}{n} q_{-2n},
\end{eqnarray}
where we have expanded the BRS current $J_{\rm B}$, the ghost and anti-ghost
fields $c$ and $b$, and the ghost number current $J_{\rm gh}$ as
\begin{eqnarray}
&& J_{\rm B}(w) = \sum_{n=-\infty}^\infty Q_n w^{-n-1},\\
&& c(w) = \sum_{n=-\infty}^\infty c_n w^{-n+1},\ \ \ 
   b(w) = \sum_{n=-\infty}^\infty b_n w^{-n-2},\\
&& J_{\rm gh}(w)= c\,b(w) = \sum_{n=-\infty}^\infty q_n w^{-n-1}.
\end{eqnarray}
Here, we have defined the ghost number current using $SL(2,R)$ normal
ordering.

It is thus found that we can reduce the cohomology of $\Q'$ to that of
the fixed level operator given in Eq.~(\ref{Eq:qQq}). In order to 
determine the cohomology of the fixed level operator,
let us consider
the replacement of the ghost and anti-ghost oscillators $c_n$ and $b_n$
by $c^{(k)}_n = c_{n+k}$ and $b^{(k)}_n=b_{n-k}$ in some 
operator $\phi$. 
We express the operator in which this replacement has been made 
by $\phi^{(k)}$.
For example, we apply
this $bc$-shift operation to the modes of the ghost number current $q_n$,
the ghost Virasoro operators $L_n^{(bc)}$, and the BRS charge $\Q$, and
then we rewrite them in terms of the
unshifted operators:
\begin{eqnarray}
q_n^{(k)}&=& q_n + k \delta_{n,0}, \\
{L_n^{(bc)}}^{(k)} &=& L_n^{(bc)}+k q_n +\half (k^2-3k) \delta_{n,0},\\
\label{Eq:BRSk}
\Q^{(k)} &=& Q_k -k^2 c_k.
\end{eqnarray}
The algebra of these $bc$-shifted operators can be calculated using 
the following commutation relations:
\begin{eqnarray}
 && [q_m,\,q_n ] = m \delta_{m+n,0}, \\
 && [L_m^{(bc)},\,L_n^{(bc)}] = (m-n)L_{m+n}^{(bc)}
    -\frac{13}{6}(m^3-m)\delta_{m+n,0}, \\
 && [L_m^{(bc)},\,q_n]=-n q_{m+n} -\frac{3}{2}m(m+1)\delta_{m+n,0}, \\
 && \{Q_m,\,Q_n\}=2mn\{Q_m,\,c_n\}, \\
 && \{Q_m,\,c_n\}=\{\Q,\,c_{m+n}\}, \\
 && \{Q_m,\,b_n\}=L_{m+n}+m q_{m+n}
    +\frac{3}{2}m(m-1)\delta_{m+n,0}, \\
 && [Q_m,\,q_n] = -Q_{m+n} + 2mn c_{m+n}.
\end{eqnarray}
It turns out that the algebra of $q_n^{(k)}$, ${L_n^{(bc)}}^{(k)}$ 
and $\Q^{(k)}$ is the same as that 
of the unshifted operators $q_n$, $L_n^{(bc)}$ and $\Q$.
In general, the algebra of $\phi^{(k)}$ is invariant 
through the $bc$-shift operation, because $c_m^{(k)}$ and $b_n^{(k)}$
satisfy the anti-commutation 
relations $\{c_m^{(k)},\,b_n^{(k)}\}=\delta_{m+n,0}$, and these
relations are the same as those before the replacement of the ghost modes.

The $SL(2,\,R)$ invariant vacuum $\ket{0}$ has the following properties
for the $bc$-shifted modes:
\begin{eqnarray}
\label{Eq:bcSLV}
&& c_n^{(k)}\ket{0} = 0,\ \ \ (n\geq 2-k)\nn
&& b_n^{(k)}\ket{0} = 0.\ \ \ (n\geq -1+k)
\end{eqnarray}
We define the $bc$-shifted vacuum $\ket{0}^{(k)}$ as
\begin{eqnarray}
 && c_n^{(k)}\ket{0}^{(k)} = 0,\ \ \ (n\geq 2)\\
&& b_n^{(k)}\ket{0}^{(k)} = 0,\ \ \ (n\geq -1)
\end{eqnarray}
From Eq.~(\ref{Eq:bcSLV}), we find that the $bc$-shifted vacuum
can be expressed in terms of the $SL(2,\,R)$ invariant vacuum:
\begin{eqnarray}
 \ket{0}^{(k)} = 
\left\{
\begin{array}{ll}
 b_{-k-1} b_{-k}\cdots b_{-2} \ket{0},& (k\geq 1)\\
 c_{k+2} c_{k+1}\cdots c_{1} \ket{0}.& (k\leq -1)
\end{array}
\right.
\end{eqnarray}

Now that the relations between the $bc$-shifted and unshifted operators
and their vacua have been established, we can determine the cohomology of the
shifted BRS charge. First, let us recall the cohomology of the
Kato-Ogawa BRS charge.\cite{rf:KO,rf:Henneaux,rf:FGZ}
\\

\noindent
{\bf Proposition 1.}
{\it Any state $\ket{\psi}$ satisfying $\Q\ket{\psi}=0$ 
can be written 
\begin{eqnarray}
\label{Eq:KO}
 \ket{\psi} = \ket{P}\otimes c_1 \ket{0} 
              +\ket{P'}\otimes c_0 c_1 \ket{0}
              +\Q\ket{\phi}.
\end{eqnarray}
Here $\ket{P}$ and $\ket{P'}$
are DDF states.\footnote{Here, we have not discussed the infrared states
with zero momenta.}
}
\\

The first term on the right-hand side of Eq.~(\ref{Eq:KO})
corresponds to the cohomology with the
additional subsidiary condition $b_0\ket{\psi} = 0$,~\cite{rf:KO} which
is the Siegel 
gauge condition in string field theory. The second term is obtained as the
additional term in case that we remove the Siegel gauge
condition.\cite{rf:Henneaux,rf:FGZ}

Since the algebra of the $bc$-shifted operators is the same as that of the
unshifted one, a similar proposition also holds if we apply
the $bc$-replacement in Proposition 1.
The BRS charge with $bc$-replacement is given by Eq.~(\ref{Eq:BRSk}).
Considering the correspondence between ($\Q^{(k)}$, $\ket{0}^{(k)}$) and
($\Q$, $\ket{0}$),
we can obtain the cohomology of $\Q^{(k)}$.
\\

\noindent
{\bf Proposition 2.}
{\it
Any state $\ket{\psi}$ satisfying $\Q^{(k)}\ket{\psi}=0$ 
can be written 
\begin{eqnarray}
 \ket{\psi} = \ket{P}\otimes c_1^{(k)}\ket{0}^{(k)}
              +\ket{P'}\otimes c_0^{(k)}c_1^{(k)}\ket{0}^{(k)}
              +\Q^{(k)}\ket{\phi}.
\end{eqnarray}
}
\\

From Eqs.~(\ref{Eq:qQq}) and (\ref{Eq:BRSk}), the new BRS charge $\Q'$
can be written as a similarity transformation of the $bc$-shifted BRS charge:
\begin{eqnarray}
\label{Eq:Qd}
 \Q' = -\frac{1}{4} e^{-q(\lambda)} \Q^{(2)} e^{q(\lambda)}.
\end{eqnarray}
Therefore, with the help of Proposition 2, any state $\ket{\psi}$ 
satisfying $\Q'\ket{\psi}=0$ can be written 
\begin{eqnarray}
\label{Eq:psi}
 \ket{\psi} = \ket{P}\otimes U b_{-2}\ket{0} 
              +\ket{P'}\otimes U \ket{0}
              +\Q' \ket{\phi},
\end{eqnarray}
where $U$ is given by
\begin{eqnarray}
 U=\exp\left(-\sum_{n=1}^\infty \frac{2}{n}q_{-2n}\right).
\end{eqnarray}
Thus, we have obtained the cohomology of the shifted BRS charge $\Q'$.

It should be noted that, for the derivation of the cohomology,
it is important that the operator $e^{q(\lambda)}$
be invertible.
Instead of Eq.~(\ref{Eq:Qd}), the new BRS charge can be written 
\begin{eqnarray}
 \Q' &=& -\frac{1}{4} e^{-q(\lambda')} \Q^{(-2)} 
 e^{q(\lambda')},
\end{eqnarray}
where $\lambda'(w)=-\log(1-w^2)$ and
\begin{eqnarray}
 q(\lambda') = 2\sum_{n=1}^\infty \frac{1}{n}q_{2n}.
\end{eqnarray}
However, we cannot reduce the cohomology of $\Q'$ to that 
of $\Q^{(-2)}$, because the operator $e^{q(\lambda')}$ has `zero modes'.
Indeed, as an example of `zero modes', we find that
\begin{eqnarray}
 e^{q(\lambda')} \times e^{-q(\lambda)}\ket{0}
=
 \exp\left(-4 \sum_{n=1}^\infty \frac{2}{n} \right)
 e^{-q(\lambda)}\ket{0} =0.
\end{eqnarray}

%%%%%%%%%%%%%%%%%%%%%%%%%%%%%%
\section{String field theory around the universal solution}
%%%%%%%%%%%%%%%%%%%%%%%%%%%%%%

The universal classical solution given in  Eq.~(\ref{Eq:solution}) has been
found from the equation of motion without gauge fixing. In fact, this
solution contains the ghost zero mode $c_0$, and therefore it is outside the
Siegel gauge. 
Therefore, 
in the context of the gauge-unfixed theory, 
a simpler argument can be given to understand the spectrum
around the solution. 

If we consider the equation
$(\Box\eta_{\mu\nu}-\partial_\mu \partial_\nu) A^\nu=0$
in an abelian
gauge theory, the solution is given 
by $A_\mu=A_\mu^{(+)}+A_\mu^{(-)}+\partial_\mu \theta$, 
where  $A_\mu^{(\pm)}$ correspond to the transverse modes 
and $\theta$ is the gauge freedom. Then, we can identify the transverse
modes as the physical degrees of freedom.
In the following, we proceed similarly. We consider the gauge-unfixed 
theory from beginning to end, where the kinetic 
operator $\Box\eta_{\mu\nu} -\partial_\mu \partial_\nu$ is 
replaced by $\Q'$; that is, 
in the gauge-unfixed theory, we
solve the equation of motion and expand the string field around
it. Then, we investigate the fluctuations in the shifted theory
without gauge-fixing and identify their physical modes up to
the gauge symmetry, not the BRS symmetry.

In open string field theory, the string field $\Psi$ is assigned the
ghost number $N_{\rm FP}=1$.\cite{rf:CSFT,rf:HIKKO}
Here, we have defined the ghost number $N_{\rm FP}$ of the $SL(2,R)$
invariant vacuum $|0\rangle$ as zero.  
Using the oscillator
representation, the string field $\Psi$ is expanded as
\begin{eqnarray}
 \ket{\Psi} = \phi(x)\,c_1\ket{0}+ A_\mu(x)\,\alpha_{-1}^\mu c_1\ket{0}
  +B(x)\,c_{-1}c_1\ket{0} + C(x) \ket{0}+\cdots.
\end{eqnarray}
The assignment $N_{\rm FP}=1$ implies that the component 
fields $\phi$, $A_\mu$, $B$ and $C$ are assigned the ghost numbers $0$, $0$,
$-1$ and $1$, respectively. In the gauge invariant theory without gauge
fixing, the ghost number of the component fields should be 
fixed to zero. In other words, the fields $B$, $C$, and so on, are forbidden.

Let us consider the perturbative spectrum in the gauge-unfixed theory
around the 
universal solution. The linearized equation 
of motion is given by
\begin{eqnarray}
\label{Eq:EOM}
 \Q'\Psi=0.
\end{eqnarray}
Using Eq.~(\ref{Eq:psi}), it is possible to find the solution of this
equation. Since the string field does not contain a component field
carrying nonzero ghost number, the solution of Eq.~(\ref{Eq:EOM}) is
given by
\begin{eqnarray}
\label{Eq:psolution}
 \Psi=\Q'\phi.
\end{eqnarray}
In the linearized approximation, the string field theory around the
solution is invariant under the gauge transformation
\begin{eqnarray}
\label{Eq:pgauge}
 \delta \Psi = \Q' \Lambda,
\end{eqnarray}
where $\Lambda$ is a gauge transformation parameter.
This symmetry is not BRS but gauge symmetry, and so both 
the parameter $\Lambda$ and its
component fields are assigned ghost number zero. 

From Eqs.~(\ref{Eq:psolution}) and (\ref{Eq:pgauge}), it follows that
all of the on-shell modes can be eliminated by the
gauge transformation in the gauge-unfixed theory.
Therefore, we conclude that the open string excitations in
the original theory disappear perturbatively after the condensation of
the string field to the universal solution.

%%%%%%%%%%%%%%%%%%%%%%%%%%%%%%
\section{Other universal solutions}
%%%%%%%%%%%%%%%%%%%%%%%%%%%%%%

We can construct other universal solutions, because 
the universal solution in Eq.~(\ref{Eq:solution}) can be constructed for 
any function $h(w)$ that makes the operator $e^{q(h)}$ singular. 
Let us consider the solution for the function
\begin{eqnarray}
 h_a^l(w)
= \log\left(1-\frac{a}{2}(-1)^l\left(w^l-(-1)^l\frac{1}{w^l}\right)^2\right),
\end{eqnarray}
where $l=1,2,\cdots$. The $l=1$ case corresponds to
the solution given previously. This function satisfies all of the conditions
discussed in \S 2, and so it yields a universal
solution. 
The function $h_a^l$ is expanded in a Laurent series as
\begin{eqnarray}
 h_a^l(w) &=& -\log(1-Z(a))^2 -\sum_{n=1}^\infty \frac{(-1)^{nl}}{n}
Z(a)^n
\left(w^{2ln}+w^{-2ln}\right).
\end{eqnarray}
From this expansion, it follows that if we 
take $a\rightarrow -1/2$ limit, the operator $e^{q(h_a^l)}$ becomes
singular. Therefore, we can obtain a non-trivial
universal solution for $h_{-1/2}^l$.

Substituting $h_{-1/2}^l$ into Eq.~(\ref{Eq:solution}), the solution is
obtained as 
\begin{eqnarray}
 \Psi_0^{(l)} = Q_\rL\left(-\frac{1}{2}+\frac{(-1)^l}4
      (w^{2l}+w^{-2l})\right)I
+C_\rL\left(l^2 w^{-2}
 \left(2-(-1)^l(w^{2l}+w^{-2l})\right) \right)I.
\end{eqnarray}
We can obtain the Fock space expression of the
universal solution as
\begin{eqnarray}
\ket{\Psi_0^{(l)}} =&&
\left[ \sum_{m=1}^\infty \frac{(-1)^m}{2\pi}
 \left(\frac{2}{2m+1}-\frac{1}{2m+1-2l}-\frac{1}{2m+1+2l}\right)
   (-Q_{-(2m+1)}+4l^2 c_{-(2m+1)})\right. \nn
&&
-\frac{1}{2\pi}\frac{8l^2}{4l^2-1} Q_{-1}
+\frac{1}{2\pi} \sum_{k=1}^l \frac{4l^2}{2k-1}\,c_1
\left.
+\frac{1}{2\pi}\left(
- \sum_{k=1}^l \frac{4l^2}{2k-1}+\frac{32l^4}{4l^2-1}
\right)c_{-1}\right] \ket{I},
\end{eqnarray}
where use has been made of the equations\cite{rf:TT2}
\begin{eqnarray}
 J_{\rm B}(w)\ket{I} &=&
  \sum_{n=1}^\infty Q_{-n}(w^n-(-1)^n w^{-n})w^{-1} \ket{I}, \nn
 c(w)\ket{I} &=& \left[
 -c_0 \frac{w-w^3}{1+w^2}+c_1 \frac{1}{1+w^2} 
 +c_{-1}\frac{1+w^2+w^4}{1+w^2}\right. \nn
&&
\left.
 +\sum_{n=2}^\infty c_{-n}(w^n-(-1)^n w^{-n})w
\right] \ket{I}.
\end{eqnarray}
If we expand the string field around this solution, the new BRS charge in
the shifted theory is given by
\begin{eqnarray}
 \tilde{Q}_{\rm B} &=& Q\left(
e^{h^l_{-1/2}}\right)
-C\left((\partial h^l_{-1/2})^2e^{h^l_{-1/2}}
\right) \\
&=&
\frac{1}{2}\Q +\frac{(-1)^l}{4}(Q_{2l}+Q_{-2l})
+2l^2 \left(c_0 -\frac{(-1)^l}{2}(c_{2l}+c_{-2l})\right).
\end{eqnarray}

Let us consider the spectrum in the shifted theory.
As in the case $l=1$, the new BRS charge can be written as a 
similarity transformation of the $bc$-shifted BRS charge.
We have
\begin{eqnarray}
 \tilde{Q}_{\rm B}=\frac{(-1)^l}{4} e^{-q(\lambda^l)}\,
 \Q^{(2l)}\, e^{q(\lambda^l)},
\end{eqnarray}
where $\lambda^l(w)=-2\log(1+(-1)^l w^{-2l})$, and $q(\lambda^l)$ is given by
\begin{eqnarray}
 q(\lambda^l)= 2\sum_{n=1}^\infty \frac{(-1)^{n(l+1)}}{n}q_{-2nl}.
\end{eqnarray}
From Proposition 2, the cohomology of the new BRS charge $\tilde{Q}_{\rm
B}\ (l\geq 2)$ can be written 
\begin{eqnarray}
 \ket{\psi}&=&
\ket{P}\otimes U_l\,b_{-2l}b_{-2l+1}\cdots b_{-2}\ket{0}\\
&& +\ket{P'}\otimes U_l\,b_{-2l+1}\cdots b_{-2}\ket{0}
+\tilde{Q}_{\rm B}\ket{\phi},
\end{eqnarray}
where $\ket{P}$ and $\ket{P'}$ denote DDF states, and $U_l$ is given by
\begin{eqnarray}
 U_l =\exp\left(-2\sum_{n=1}^\infty \frac{(-1)^{n(l+1)}}{n}q_{-2nl}\right).
\end{eqnarray}
Thus, the non-trivial cohomology is assigned to the ghost numbers $-2l+1$
and $-2l+2$.
Therefore, again applying the argument used in the case $l=1$, we
conclude that 
there is no physical spectrum perturbatively in the string field theory
around the solution.

\section{Summary and discussion}

We have obtained the cohomology for the new BRS charge around the universal
solution in CSFT. The non-trivial states belong to the ghost number $0$
and $-1$ sectors. Consequently, the shifted theory around the solution
contains no physical spectrum perturbatively. We have found other
possible solutions in CSFT. Around these solutions too,
the shifted theories have no physical spectrum.

We believe that the universal solutions correspond to the tachyon vacuum
solution, because they have universal expressions and there is no
physical spectrum around them. However, the universal solutions have
several possible expressions. If these  
solutions represent the tachyon condensation,
they must be equivalent
up to gauge transformations, and the shifted theories must be
related by string field redefinitions.

Let us consider the string field redefinition generated by the operator
$K_n=L_n-(-1)^n L_{-n}\ (n\geq 1)$. This field redefinition preserves the
three string vertex, and it only changes the kinetic operator
$\tilde{Q}_{\rm B}$ to $e^K \tilde{Q}_{\rm B} e^{-K}$.\cite{rf:SCSFT,rf:RSZ}
Since the operators $Q_n$ and $c_n$ have the algebra
\begin{eqnarray}
\left[K_m, \, Q_n \right]
&=& -n Q_{n+m}+ (-1)^m n Q_{n-m}, \\
\left[
K_m,\, c_n \right]
 &=& -(2m+n) c_{n+m} -(-1)^m (2m-n) c_{n-m},
\end{eqnarray}
the transform of the new BRS charge
remains of the form $Q(f)+C(g)$, for some functions $f(w)$ and $g(w)$.
As pointed out in \S 2, the BRS charge with this form must be
written as $Q(e^h)-C((\partial h)^2 e^h)$, because of its nilpotency.
Therefore, it is sufficient to consider the ghost kinetic parts in order
to determine the number of inequivalent classes of the new BRS charge through
the string field redefinition $\Psi'=e^K \Psi$.

According to Ref.~\citen{rf:RSZ}, with this field redefinition, the ghost
kinetic terms $a_0c_0+\sum_{n\geq 1} 
a_n (c_n+(-1)^n c_{-n})$ can be classified into at least two classes: those
which annihilate the identity string field and those which do not. For
all our solutions, the ghost parts annihilate the
identity string field.\cite{rf:TT,rf:Schnabl2}
Therefore, there is no contradiction at present if the shifted theories
can be transformed into each other through field redefinitions.
The feasibility of such a procedure requires further investigation.

At present, there are many important issues that remain incompletely
understood.
In this paper, we did not discuss the potential height at our solutions, but
determining this is an important problem in order to conclude whether or
not they correspond to the tachyon vacuum. 
Unfortunately, it is difficult to evaluate the action at the solution
that is constructed using the identity string field, because we often encounter
disastrous divergences in the computation of quantities of 
the form $\langle I |\dots |
I\rangle$.  We should find a consistent regularization in order to treat
the identity string field correctly in these calculations.  

We should comment on the relation between our result and that in
Ref.~\citen{rf:EFHM}. In that work, it is shown
in terms of the level truncation scheme
that the cohomology of the new BRS
charge around the tachyon vacuum is trivial.
This would seem to contradict our result if the universal
solutions represent the tachyon condensation, because our solutions
yield a non-trivial cohomology.
We have not yet understood this apparent contradiction.

As is well known, vacuum string field theory (VSFT)
\cite{rf:VSFT1,rf:VSFT2} is 
another approach to investigate the tachyon condensation, and there is
some evidence that VSFT represents the theory around the tachyon
vacuum. \cite{rf:Okawa} 
It would be interesting to examine the relations between our
solutions in the context of CSFT and VSFT. 

Around the tachyon vacuum, there would be closed string excitations
instead of open ones. 
It is a future problem to investigate how to treat closed string around
our solutions in the context of CSFT. It might be necessary to fix the
gauge appropriately and perform a second quantization.

\section*{Acknowledgements}
We would like to thank H.~Hata, M.~Kato and T.~Kugo for valuable
discussions and comments. 
I.~K. wishes to express his gratitude to Y.~Matsuo and K.~Ohmori for helpful discussions and encouragement.
I.~K. is supported in part by JSPS Research Fellowships for Young Scientists.

%\appendix
%\section{First Appendix} %Empty argument \section{} yields `Appendix'. 
%
%\section{Second Appendix}

%\newpage

\end{document}